\documentclass[aps,prd,onecolumn,groupedaddress,showpacs,nofootinbib,amssymb]{revtex4}
\usepackage[dvips]{graphicx}
\usepackage{amssymb}
\usepackage{amsmath}
\usepackage{graphicx,,color}
\usepackage{amsfonts}
\usepackage{bm}
\usepackage{cancel}
\usepackage{comment}

\newcommand\be{\begin{equation}}
\newcommand\ee{\end{equation}}

\allowdisplaybreaks[4]

\begin{document}

\title{$R^2$ Inflation Revisited and Dark Energy Corrections}
\author{S.D. Odintsov,$^{1,2}$}
\email{odintsov@ice.cat}\author{V.K.
Oikonomou,$^{3,4}$}\email{v.k.oikonomou1979@gmail.com,voikonomou@auth.gr}
\affiliation{$^{1)}$ ICREA, Passeig Luis Companys, 23, 08010 Barcelona, Spain\\
$^{2)}$ Institute of Space Sciences (IEEC-CSIC) C. Can Magrans
s/n,
08193 Barcelona, Spain\\
$^{3)}$Department of Physics, Aristotle University of Thessaloniki, Thessaloniki 54124, Greece\\
$^{4)}$ Laboratory for Theoretical Cosmology, Tomsk State
University of Control Systems and Radioelectronics  (TUSUR),
634050 Tomsk, Russia}

\tolerance=5000

\begin{abstract}
The vacuum $R^2$ model is known to generate a quasi-de Sitter
evolution for inflation, using solely the slow-roll assumptions.
Using standard reconstruction techniques, we demonstrate that the
$f(R)$ gravity which actually realizes the quasi-de Sitter
evolution is not simply the $R^2$ model but a deformed $R^2$ model
which contains extra terms in addition to the $R^2$ model. We
analyze in detail the inflationary dynamics of the deformed $R^2$
model and we demonstrate that the predictions are quite close to
the ones of the pure $R^2$ model, regardless the values of the
free parameters. Basically the deformed $R^2$ model is also a
single parameter inflationary model, exactly like the ordinary
$R^2$ model. In contrast to the early-time era, where the deformed
$R^2$ model is quite similar to the $R^2$ model, at late times,
the phenomenological picture is different. The deformed $R^2$
model describes stronger gravity compared to the ordinary $R^2$
with an effectively smaller effective Planck mass. We propose the
addition of an early dark energy term which does not affect at all
the inflationary era, but takes over the control of the late-time
dynamics at late times. We study in some detail the predicted dark
energy era evolution, and we demonstrate that the dark energy
corrected deformed $R^2$ model can describe a viable dark energy
era, compatible with the Planck constraints on cosmological
parameters. Furthermore the model is distinct from the
$\Lambda$-Cold-Dark-Matter model, but shows a qualitatively
similar behavior.
\end{abstract}


\maketitle

\section{Introduction}

We are nearing the stage four epoch of Cosmic  Microwave
Background (CMB) experiments, and all the scientific research
related to cosmology are focused on finding the $B$-modes of
inflation. In general, these curl modes of the CMB can be
generated in two ways, firstly from the gravitational lensing
conversion of $E$-modes to $B$-modes at small angular scales or
large multipoles of the CMB, or by the tensor perturbations at low
multipoles of the CMB or at large angular scales. The latter is
the main focus of the stage-4 CMB experiments, such as CMB-S4
\cite{CMB-S4:2016ple} and the Simons Observatory
\cite{SimonsObservatory:2019qwx}. In addition to these stage four
CMB experiments, the scientific community of astrophysicists and
cosmologists are eagerly anticipating the results of several other
scientific experiments and space observatories, which may reveal
the existence of primordial gravitational waves, among other
things. Future collaborations and experiments like the Einstein
Telescope \cite{Hild:2010id}, the LISA Space-borne Laser
Interferometer Space Antenna \cite{Baker:2019nia,Smith:2019wny}
the BBO \cite{Crowder:2005nr,Smith:2016jqs}, DECIGO
\cite{Seto:2001qf,Kawamura:2020pcg} and finally the SKA (Square
Kilometer Array) Pulsar Timing Arrays \cite{Bull:2018lat} are
eagerly expected from theoretical cosmologists and
astrophysicists, since all the aforementioned collaborations will
shed light to the most mysterious era of our Universe, the
early-time post-Planck era. Indeed, inflation and reheating and
even the early stages of the radiation domination era will
possibly reveal their secrets via the primordial gravitational
wave spectrum. In some sense, high energy physicists, theoretical
cosmologists and theoretical astrophysicists will focus their
interest on the CMB stage 4 experiments and on the space
collaborations capturing primordial gravitational waves. The
future of theoretical particle physics seems to be in the sky
eventually.

In view of these future and nearing experiments, many theoretical
models will be scrutinized and stress tests will verify their
validity and the limits of their validity. Inflation
\cite{inflation1,inflation2,inflation3,inflation4} is one of the
most successful and elegant theoretical descriptions of the
post-Planck early time, since it explains in a rigid way most of
the shortcomings of the standard Big Bang cosmology. However, to
date, no sign of the $B$-mode polarization in the CMB has ever
been observed, nor a sign of a stochastic gravitational wave
background has ever been observed. It is conceivable that if a
direct signal of $B$-modes in the CMB is observed
\cite{Kamionkowski:2015yta}, or if a stochastic gravitational wave
background is observed, this would signify directly that inflation
indeed took place.

With all the upcoming experiments which are expected to stir up
things in theoretical cosmology, many models of inflation might be
in peril to become non-viable. In most cases, inflationary models
rely on single scalar fields, minimally or non-minimally coupled.
However, another promising approach which can realize a viable
inflationary era without relying to scalar fields, or at least not
relying solely on scalar fields, is modified gravity in its
various forms
\cite{reviews1,reviews2,reviews3,reviews4,reviews5,reviews6}.
There are many kinds of modified gravities that can successfully
describe inflation, such as $f(R)$ gravity
\cite{Nojiri:2003ft,Capozziello:2005ku,Capozziello:2004vh,Capozziello:2018ddp,Hwang:2001pu,Cognola:2005de,Nojiri:2006gh,Song:2006ej,Capozziello:2008qc,Bean:2006up,Capozziello:2012ie,Faulkner:2006ub,Olmo:2006eh,Sawicki:2007tf,Faraoni:2007yn,Carloni:2007yv,
Nojiri:2007as,Capozziello:2007ms,Deruelle:2007pt,Appleby:2008tv,Dunsby:2010wg,Odintsov:2020nwm,Odintsov:2019mlf,Odintsov:2019evb,Oikonomou:2020oex,Oikonomou:2020qah},
Gauss-Bonnet gravity
\cite{Nojiri:2005vv,Nojiri:2005jg,Lidsey:2003sj,Carter:2005fu,Cognola:2006eg,Leith:2007bu,Li:2007jm,Bamba:2010wfw,Odintsov:2020sqy,Kanti:2015dra,
Oikonomou:2015qha,Kanti:2015pda,Oikonomou:2021kql} and so on, and
in some cases a unified description of the inflationary era with
the dark energy era can also be described even with the same
model, as in the pioneer work \cite{Nojiri:2003ft}, see also
\cite{Nojiri:2006gh,Nojiri:2007as,Appleby:2008tv,Odintsov:2019evb,Oikonomou:2020oex,Oikonomou:2020qah}
for later developments. In this work we shall consider
deformations of $R^2$ gravity
\cite{Starobinsky:1980te,Bezrukov:2007ep} caused by demanding a
quasi de Sitter evolution during the early-time era, the widely
known in the literature as Starobinsky inflation. The reason for
this is simple, it is known in the literature that if we solve the
$R^2$ gravity equations of motion using the slow-roll conditions,
one obtains a quasi-de Sitter evolution. However, in the converse
way, if we demand a quasi-de Sitter evolution to be realized by
vacuum $f(R)$ gravity, by using standard reconstruction
techniques, we will show that the resulting $f(R)$ gravity
contains the $R^2$ model, but there are also other terms in the
final form of the quasi-de Sitter realizing $f(R)$ gravity. We
thoroughly investigate the effect of the extra terms in the
inflationary Lagrangian, on the inflationary dynamics of the
model, using only the slow-roll condition. As we show, the
resulting model is nothing but a slight deformation of the $R^2$
model, the dynamics are very similar and almost identical.
However, at late times, the extra terms might affect the dynamics
significantly. Specifically, in the small curvature limit, which
describes the late-time era, the effective Planck mass is
effectively smaller compared to the early time era, thus in some
sense, effectively gravity becomes stronger at late times. We also
propose the addition of an early dark energy term in the full
deformed $R^2$ Lagrangian, which does not affect at all the
inflationary era, but strongly affects the late-time era. As we
show, the dark energy deformed $R^2$ model provides a viable
late-time phenomenology at late times, and serves as a viable
deformation of the $\Lambda$-Cold-Dark-Matter ($\Lambda$CDM)
model.

This paper is organized as follows: In section II we discuss the
essential features of the quasi-de Sitter evolution and how can
this evolution be realized by a vacuum $f(R)$ gravity. We also
calculate in the same section the form of the $f(R)$ gravity and
that the resulting form is a deformation of the $R^2$ model. In
section III we study in detail the inflationary phenomenology of
this model and discuss several limiting cases of the deformed
$R^2$ model. In section IV we propose an early dark energy term
which does not affect inflation at all, but does affect the
late-time era significantly, and we show that the dark energy
corrected deformed $R^2$ model can produce a viable dark energy
era and mimic to some extent the $\Lambda$CDM model. Finally, the
conclusions follow in the end of the paper.

Before starting, let us mention that he geometric background which
will be assumed, is that of a flat Friedmann-Robertson-Walker
(FRW) background of the form,
\begin{equation}
\label{JGRG14} ds^2 = - dt^2 + a(t)^2 \sum_{i=1,2,3}
\left(dx^i\right)^2\, ,
\end{equation}
where $a(t)$ being as usual the scale factor. We shall also adopt
the natural units physical system of units.

\section{Quasi-de Sitter Evolution and Realization from $f(R)$ Gravity}

\begin{figure}[h!]
\centering
\includegraphics[width=16pc]{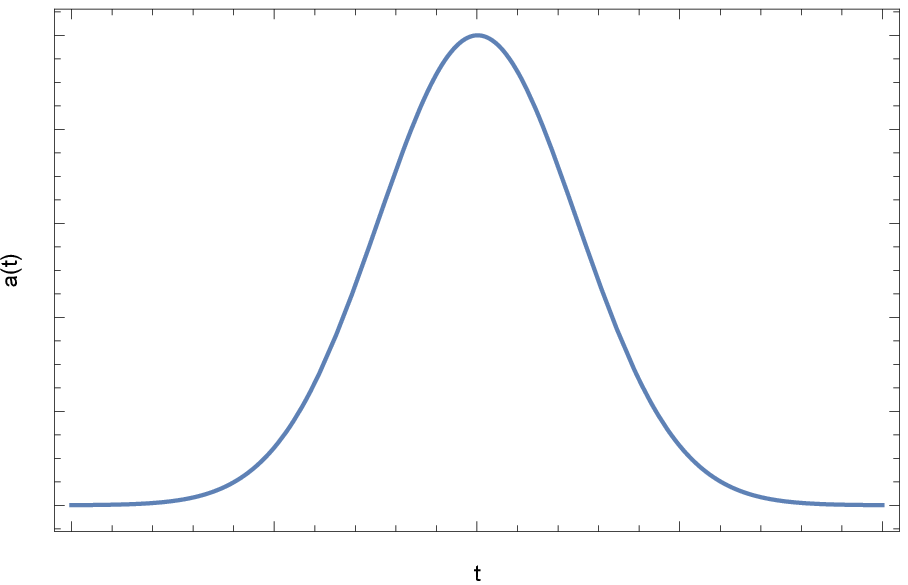}
\includegraphics[width=16pc]{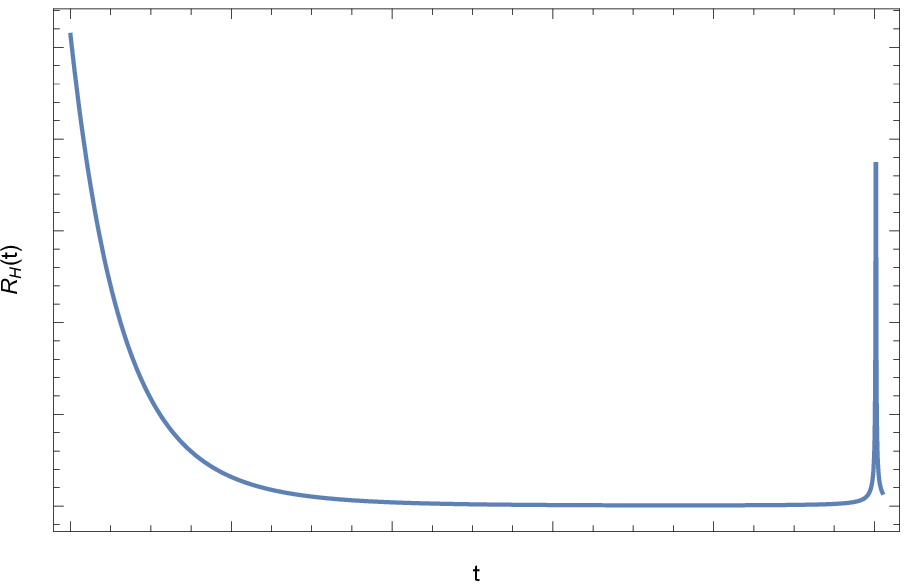}
\includegraphics[width=16pc]{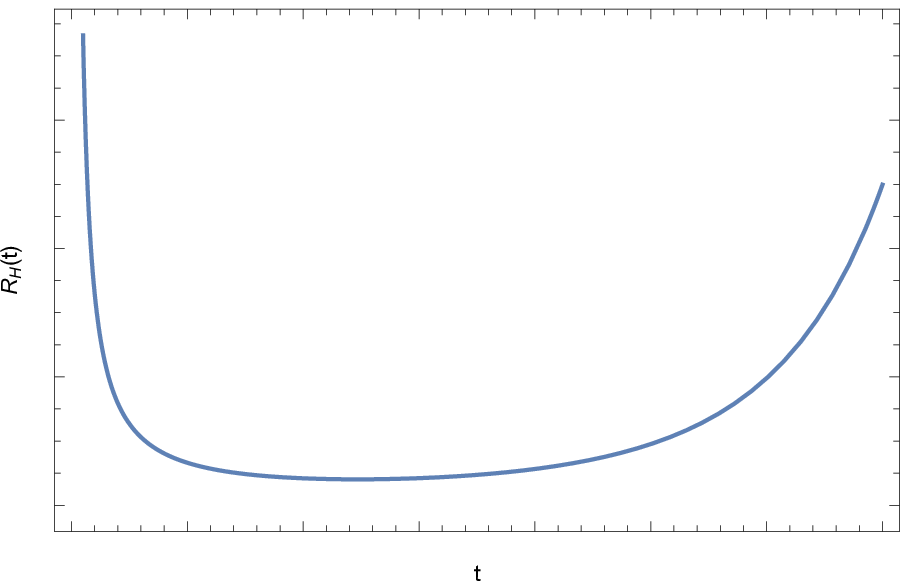}
\caption{The scale factor of the quasi-de Sitter evolution (left
plot) and the corresponding Hubble radius (right plot) and a
close-up of the Hubble radius (bottom plot).} \label{plot1}
\end{figure}

The inflationary era is basically realized in general relativity
by a de Sitter or a quasi-de Sitter evolution at least. It is
quite difficult to capture the entire evolution of the Universe
during the post-Planck era, however, we have phenomenological and
theoretical hints on how the Universe evolves during the various
cosmological eras. One hint is that during an accelerating era,
the Universe must evolve for a large number of $e$-foldings in a
quasi-de Sitter way. We need to stress that this quasi-de Sitter
era is a transient era and not a permanent state. The physics of
the effective inflationary Lagrangian will stop the quasi-de
Sitter era at some point and the Universe will enter the
mysterious reheating era, in which the equation of state (EoS)
will be described by radiation or some variant form in alternative
kination scenarios. Here we shall focus on the quasi-de Sitter
phase and we shall analyze in detail the physics of it. We shall
reveal interesting features of the standard $R^2$ model, which
elevate the role of the standard $R^2$ model to one of the most
elegant descriptions of the early Universe, to date at least. To
start off, consider the following quasi-de Sitter evolution,
\begin{equation}\label{quasidesitter}
H=\mathcal{H}_0-\frac{M^2}{6}(t-t_i)\, ,
\end{equation}
where $\mathcal{H}_0$ is basically the scale of inflation and has
mass dimensions in natural units $\mathcal{H}_0=[m]$, $M$ is a
parameter which deforms the de-Sitter to a quasi-de Sitter
evolution, which plays an important role in the inflationary
evolution, and $t_i$ is the time instance on which the quasi-de
Sitter evolution commences. From Eq. (\ref{quasidesitter}) we
easily obtain the scale factor of the quasi-de Sitter evolution,
which is,
\begin{equation}\label{scalefactorquasidesitter}
a(t)=a_0e^{\mathcal{H}_0 t-\frac{1}{12} M^2 t^2}\, ,
\end{equation}
and the normalization factor can be taken be unity in order for
the comoving scales to be identical with real physical scales
during inflation, but we leave this as it is. Let us study the
behavior of the quasi-de Sitter evolution (\ref{quasidesitter}) in
order to further understand this evolution patch of our Universe
and to better understand this inflationary evolution. In Fig.
\ref{plot1} we present the behavior of the scale factor as a
function of the cosmic time (left plot) and the Hubble radius
$R_H=\frac{1}{a(t)H(t)}$ as a function of the cosmic time (right
plot). As it can be seen, the scale factor grows exponentially for
a sufficient amount of time, however the quasi-de Sitter patch of
the Universe starts to drop after some point in time. It is clear
that this time instance must be the one for which inflation ends.
The Hubble radius accordingly drops until some point where it
starts to grow again. Basically the point at which the Hubble
radius reaches a minimum, basically indicates the end of the
inflationary era.
We can calculate analytically this point by finding the critical
points of the Hubble radius upon solving
$\dot{R}_H(t)=-\frac{\ddot{a}(t)}{\dot{a}(t)^2}=0$, which yields,
\begin{equation}\label{grexit}
t_{f}=\frac{6 \mathcal{H}_0-\sqrt{6} M}{M^2}\, ,
\end{equation}
and is basically the time instance for which $\ddot{a}(t)=0$. At
this time instance the acceleration era ends, and the Universe
starts to decelerate again. Remarkably, if one calculates, for the
same scale factor the solution to the equation
$\epsilon_1=-\frac{\dot{H}}{H^2}=1$, one obtains the solution
(\ref{grexit}), and recall that the condition for which the first
slow-roll index $\epsilon_1$ becomes of the order $\sim
\mathcal{O}(1)$, namely, $\epsilon_1=-\frac{\dot{H}}{H^2}=1$
indicates the time instance for which inflation ends. This is a
remarkable feature of the quasi-de Sitter evolution
(\ref{quasidesitter}) which describes in an accurate way a smooth
and phenomenologically appealing inflationary era, since the time
instance for which inflation ends is exactly the time instance for
which the Hubble radius starts to increase. This behavior should
be obtained from any viable inflationary model and it describes an
ideal and desirable inflationary evolution.

Let us now demonstrate how the quasi-de Sitter cosmology
(\ref{quasidesitter}) can be realized by $f(R)$ gravity. We
consider a vacuum $f(R)$ gravity theory with action,
\begin{equation}\label{action1dse}
\mathcal{S}=\frac{1}{2\kappa^2}\int \mathrm{d}^4x\sqrt{-g}f(R),
\end{equation}
with $\kappa^2$ being $\kappa^2=8\pi G=\frac{1}{M_p^2}$ and with
$M_p$denoting the reduced Planck mass. In the metric formalism,
the field equations can be obtained by varying the gravitational
action with respect to the metric tensor, so we obtain,
\begin{equation}\label{eqnmotion}
f_R(R)R_{\mu \nu}(g)-\frac{1}{2}f(R)g_{\mu
\nu}-\nabla_{\mu}\nabla_{\nu}f_R(R)+g_{\mu \nu}\square f_R(R)=0\,
,
\end{equation}
with $f_R=\frac{\mathrm{d}f}{\mathrm{d}R}$. We can rewrite
(\ref{eqnmotion}) as follows,
\begin{align}\label{modifiedeinsteineqns}
R_{\mu \nu}-\frac{1}{2}Rg_{\mu
\nu}=\frac{\kappa^2}{f_R(R)}\Big{(}T_{\mu
\nu}+\frac{1}{\kappa^2}\Big{(}\frac{f(R)-Rf_R(R)}{2}g_{\mu
\nu}+\nabla_{\mu}\nabla_{\nu}f_R(R)-g_{\mu \nu}\square
f_R(R)\Big{)}\Big{)}\, .
\end{align}
For the FRW metric (\ref{JGRG14}), the field equations take the
form,
\begin{align}
\label{JGRG15} 0 =& -\frac{f(R)}{2} + 3\left(H^2 + \dot H\right)
f_R(R) - 18 \left( 4H^2 \dot H + H \ddot H\right) f_{RR}(R)\, ,\\
\label{Cr4b} 0 =& \frac{f(R)}{2} - \left(\dot H +
3H^2\right)f_R(R) + 6 \left( 8H^2 \dot H + 4 {\dot H}^2 + 6 H
\ddot H + \dddot H\right) f_{RR}(R) + 36\left( 4H\dot H + \ddot
H\right)^2 f_{RRR} \, ,
\end{align}
with $f_{RR}=\frac{\mathrm{d}^2f}{\mathrm{d}R^2}$, and
$f_{RRR}=\frac{\mathrm{d}^3f}{\mathrm{d}R^3}$, with $H$ being the
Hubble rate $H=\dot a/a$ and the Ricci scalar for the FRW metric
is $R=12H^2 + 6\dot H$.

We can find which $f(R)$ gravity can realize the quasi-de Sitter
evolution (\ref{quasidesitter}) by using the reconstruction
technique developed in  \cite{Nojiri:2009kx}, which is based on
using the $e$-foldings number instead of the cosmic time, defined
in terms of the scale factor as,
\begin{equation}\label{efoldpoar}
e^{-N}=\frac{a_0}{a}.
\end{equation}
In terms of $N$, the Friedmann equation (\ref{JGRG15}) is written
as,
\begin{equation}
\label{newfrw1} -18\left [ 4H^3(N)H'(N)+H^2(N)(H')^2+H^3(N)H''(N)
\right ]f_{RR}+3\left [H^2(N)+H(N)H'(N)
\right]f_R-\frac{f(R)}{2}=0,
\end{equation}
where the prime denotes differentiation with respect to the
$e$-foldings number. We introduce the function $G(N)=H^2(N)$, and
therefore the Ricci scalar reads,
\begin{equation}\label{riccinrelat}
R=3G'(N)+12G(N)\, ,
\end{equation}
and from this one can easily obtain the function $N(R)$. For the
quasi-de Sitter evolution (\ref{quasidesitter}), we have,
\begin{equation}\label{gnfunction}
G(N)=\mathcal{H}_0^2-\frac{M^2 N}{3}\, .
\end{equation}
Upon combining Eqs. (\ref{riccinrelat}) and (\ref{gnfunction}), we
get the $e$-foldings number $N$ as a function of the Ricci scalar,
\begin{equation}\label{efoldr}
N=\frac{12 \mathcal{H}_0^2-M^2-R}{4 M^2}\, .
\end{equation}
The Friedmann equation can be written in terms of the function
$G(N)$ as follows,
 \begin{equation}
\label{newfrw1modfrom} -9G(N(R))\left[ 4G'(N(R))+G''(N(R))
\right]F''(R) +\left[3G(N)+\frac{3}{2}G'(N(R))
\right]F'(R)-\frac{F(R)}{2}=0,
\end{equation}
with $G'(N)=\mathrm{d}G(N)/\mathrm{d}N$ and
$G''(N)=\mathrm{d}^2G(N)/\mathrm{d}N^2$. By using Eq.
(\ref{efoldr}), the Friedmann equation reads,
\begin{align}
\label{bigdiffgeneral1} & M^2
\left(M^2+R\right)\frac{\mathrm{d}^2f(R)}{\mathrm{d}R^2}
+\frac{1}{4}
\left(R-M^2\right)R\frac{\mathrm{d}f(R)}{\mathrm{d}R}-\frac{f(R)}{2}=0,
\end{align}
which can be solved and yields the $f(R)$ gravity which can
realize the quasi-de Sitter evolution (\ref{quasidesitter}), which
is,
\begin{equation}\label{frformprev}
f(R)=\frac{\mathcal{C}_1 \left(M^4+6 M^2
R+R^2\right)}{M^4}-\frac{\mathcal{C}_2 \left(M^4+6 M^2
R+R^2\right) \left(\sqrt[4]{e} \sqrt{\pi }
\mathrm{Erf}\left(\frac{\sqrt{M^2+R}}{2 M}\right)+\frac{2 M
e^{-\frac{R}{4 M^2}} \left(3 M^2+R\right) \sqrt{M^2+R}}{M^4+6 M^2
R+R^2}\right)}{32 M^9}\, ,
\end{equation}
where $\mathcal{C}_1$ and $\mathcal{C}_2$ are integration
constants. The standard coupling of the Einstein-Hilbert term $R$
can be obtained by choosing $\mathcal{C}_1=\frac{M^2}{6}$, so we
have,
\begin{equation}\label{frfinal}
f(R)=R+\frac{R^2}{6 M^2}+\frac{M^2}{6}-\frac{\mathcal{C}_2
\left(M^4+6 M^2 R+R^2\right) \left(\sqrt[4]{e} \sqrt{\pi }
\mathrm{Erf}\left(\frac{\sqrt{M^2+R}}{2 M}\right)+\frac{2 M
e^{-\frac{R}{4 M^2}} \left(3 M^2+R\right) \sqrt{M^2+R}}{M^4+6 M^2
R+R^2}\right)}{32 M^9}\, ,
\end{equation}
where the constant of integration $\mathcal{C}_2$ has mass
dimensions $[\mathcal{C}_2]=[m]^7$. As it can be seen, the $f(R)$
gravity contains the standard $R^2$ model $R+\frac{R^2}{6 M^2}$
among other alternative terms. The standard $R^2$ gravity in the
slow-roll approximation realizes the quasi-de Sitter evolution,
and using the reconstruction technique, without assuming for the
moment the slow-roll condition, we demonstrated that the quasi-de
Sitter evolution is not realized by the $R^2$ model solely, but
from the $f(R)$ gravity of Eq. (\ref{frfinal}). In the next
section, we shall analyze in detail the inflationary predictions
of the model (\ref{frfinal}).

\section{Inflationary Phenomenology of the quasi-de Sitter Realizing Extended $R^2$ Gravity}

We shall be interested in analyzing in some detail the
inflationary phenomenology of the $f(R)$ gravity appearing in Eq.
(\ref{frfinal}) which is basically a deformation of the $R^2$
model, under the assumption of slow-roll dynamics,
\begin{equation}\label{slowrollconditionshubble}
\ddot{H}\ll H\dot{H},\,\,\, \frac{\dot{H}}{H^2}\ll 1\, .
\end{equation}
The slow-roll indices, namely $\epsilon_1$ ,$\epsilon_2$,
$\epsilon_3$, $\epsilon_4$, basically quantify the inflationary
era dynamics and for the case of $f(R)$ gravity, these read
\cite{Hwang:2005hb,reviews1},
\begin{equation}
\label{restofparametersfr}\epsilon_1=-\frac{\dot{H}}{H^2}, \quad
\epsilon_2=0\, ,\quad \epsilon_3= \frac{\dot{f}_R}{2Hf_R}\, ,\quad
\epsilon_4=\frac{\ddot{f}_R}{H\dot{f}_R}\,
 .
\end{equation}
Assuming that $\epsilon_i\ll 1$, $i=1,3,4$ and the observational
indices of inflation, namely the spectral index of the primordial
scalar perturbations, the tensor-to-scalar ratio and the spectral
index of the tensor perturbations, expressed in terms of the
slow-roll indices are \cite{reviews1,Hwang:2005hb,Kaiser:1994vs},
\begin{equation}
\label{epsilonall} n_s=
1-\frac{4\epsilon_1-2\epsilon_3+2\epsilon_4}{1-\epsilon_1},\quad
r=48\frac{\epsilon_3^2}{(1+\epsilon_3)^2}\, \quad
n_T=-2(\epsilon_1+\epsilon_3)\, .
\end{equation}
From the Raychaudhuri equation, in the case of vacuum $f(R)$
gravity we get the following relation,
\begin{equation}\label{approx1}
\epsilon_1=-\epsilon_3(1-\epsilon_4)\, .
\end{equation}
Now with regard to the slow-roll index $\epsilon_4$, we shall
express it in terms of the first slow-roll index. After some
algebra we have,
\begin{equation}\label{epsilon41}
\epsilon_4=\frac{\ddot{f}_R}{H\dot{f}_R}=\frac{\frac{d}{d
t}\left(f_{RR}\dot{R}\right)}{Hf_{RR}\dot{R}}=\frac{f_{RRR}\dot{R}^2+f_{RR}\frac{d
(\dot{R})}{d t}}{Hf_{RR}\dot{R}}\, ,
\end{equation}
and by using the slow-roll assumption, $\dot{R}$ is approximately
equal to,
\begin{equation}\label{rdot}
\dot{R}=24\dot{H}H+6\ddot{H}\simeq 24H\dot{H}\simeq
-24H^3\epsilon_1\, .
\end{equation}
where we used the slow-roll approximation condition $\ddot{H}\ll H
\dot{H}$. Then by combining Eqs. (\ref{rdot}) and
(\ref{epsilon41}) after some algebra we obtain,
\begin{equation}\label{epsilon4final}
\epsilon_4\simeq -\frac{24
F_{RRR}H^2}{F_{RR}}\epsilon_1-3\epsilon_1+\frac{\dot{\epsilon}_1}{H\epsilon_1}\,
,
\end{equation}
and since $\dot{\epsilon}_1$ is,
\begin{equation}\label{epsilon1newfiles}
\dot{\epsilon}_1=-\frac{\ddot{H}H^2-2\dot{H}^2H}{H^4}=-\frac{\ddot{H}}{H^2}+\frac{2\dot{H}^2}{H^3}\simeq
2H \epsilon_1^2\, ,
\end{equation}
we obtain the final expression for the slow-roll index
$\epsilon_4$ which is,
\begin{equation}\label{finalapproxepsilon4}
\epsilon_4\simeq -\frac{24
F_{RRR}H^2}{F_{RR}}\epsilon_1-\epsilon_1\, .
\end{equation}
Let us apply the formalism for the specific deformation of the
$R^2$ model in Eq. (\ref{frfinal}). For convenience, let us
introduce the parameter $x=-\frac{48 F_{RRR}H^2}{F_{RR}}$, so that
$\epsilon_4$ is written in terms of it,
\begin{equation}\label{espilonfourfin}
\epsilon_4\simeq \frac{x}{2}\epsilon_1-\epsilon_1\, .
\end{equation}
The parameter $x$ basically quantifies the deformation of the
model (\ref{frfinal}) compared to the $R^2$ model.

\begin{figure}[h!]
\centering
\includegraphics[width=22pc]{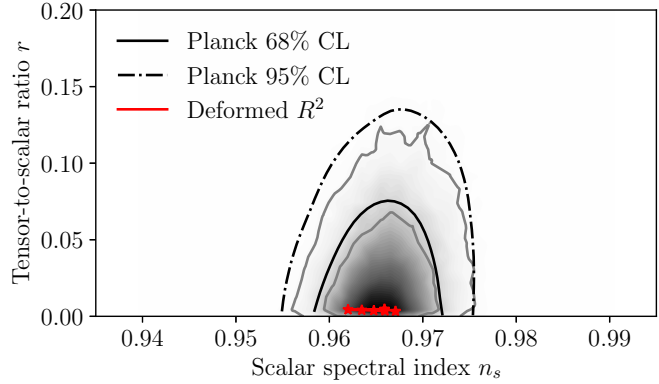}
\caption{The deformed $R^2$ model joined $n_s-r$ plots in view of
the Planck 2018 data for $N=[50,60]$.} \label{plotplanck}
\end{figure}

Let us calculate in detail the slow-roll indices for the case at
hand, starting with $\epsilon_1$ which is,
\begin{equation}\label{firstslowroll}
\epsilon_1=\frac{6 M^2}{\left(M^2 t-6 \mathcal{H}_0\right)^2}\, ,
\end{equation}
and $\epsilon_3$ can easily be found from Eq. (\ref{approx1}). For
$\epsilon_4$ it suffices to calculate the parameter $x$ defined
below Eq. (\ref{finalapproxepsilon4}), which reads,
\begin{equation}\label{xdeformedr2model}
x=-48\frac{36 \sqrt{3} \sqrt[4]{e} \mathcal{C}_2 M^3}{\left(M^2
t-6 \mathcal{H}_0\right)^2 \mathcal{S}_1}\, ,
\end{equation}
where $\mathcal{S}_1$ is defined as,
\begin{align}\label{mathcals1new}
 \mathcal{S}_1=&3 \sqrt{\pi } \mathcal{C}_2 e^{\frac{3
\left(\mathcal{H}_0-\frac{M^2 t}{6}\right)^2}{M^2}+\frac{1}{4}}
\sqrt{\left(M^2 t-6 \mathcal{H}_0\right)^2}\times\\ \notag &
\mathrm{Erf}\left(\frac{\sqrt{3}
\sqrt{\left(\mathcal{H}_0-\frac{M^2 t}{6}\right)^2}}{M}\right)+6
\sqrt[4]{e} \sqrt{3} \mathcal{C}_2 M-16 M^7 e^{\frac{3
\left(\mathcal{H}_0-\frac{M^2 t}{6}\right)^2}{M^2}}
\sqrt{\left(M^2 t-6 \mathcal{H}_0\right)^2}
\end{align}
By solving $\epsilon_1=\mathcal{O}(1)$ we may obtain the time
instance that inflation ends, which is $t_f=\frac{6
\mathcal{H}_0-\sqrt{6} M}{M^2}$, and by solving the equation
$N=\int_{t_i}^{t_f}H\mathrm{d}t$ for the quasi-de Sitter evolution
(\ref{quasidesitter}) with respect to $t_i$, we may obtain the
latter which is the time instance at the first horizon crossing.
At this time instance one needs to evaluate the slow-roll indices
and the corresponding observational indices. The first horizon
crossing time instance reads, $t_i=\frac{6
\left(\mathcal{H}_0-\frac{M \sqrt{2 N+1}}{\sqrt{6}}\right)}{M^2}$.
Evaluating the spectral index of scalar perturbations at $t=t_i$
for $60$ $e$-foldings we get the following approximate expression,
\begin{align}\label{spectralindexapproxexpre60}
 n_s\simeq &\frac{9 \mathcal{C}_2^2 \left(157036 e^{121} \pi
\mathrm{Erf}\left(\frac{11}{\sqrt{2}}\right)^2+28189 e^{121/2}
\sqrt{2 \pi }
\mathrm{Erf}\left(\frac{11}{\sqrt{2}}\right)+2530\right)}{121
\mathcal{S}_2 \mathcal{S}_3 }\\ \notag & -\frac{48 e^{241/4}
\mathcal{C}_2 \left(314072 e^{121/2} \sqrt{\pi }
\mathrm{Erf}\left(\frac{11}{\sqrt{2}}\right)+28189 \sqrt{2}\right)
M^7+40201216 e^{241/2} M^{14}}{121\mathcal{S}_2 \mathcal{S}_3}\, ,
\end{align}
where $\mathcal{S}_1$ and $\mathcal{S}_2$ are defined as follows,
\begin{align}\label{mathcals1s2}
\mathcal{S}_2=&\left(3 \mathcal{C}_2 \left(11 e^{121/2} \sqrt{\pi
} \mathrm{Erf}\left(\frac{11}{\sqrt{2}}\right)+\sqrt{2}\right)-176
e^{241/4} M^7\right)\, ,\\ \notag & \mathcal{S}_3=\left(366
e^{121/2} \sqrt{\pi } \mathcal{C}_2
\mathrm{Erf}\left(\frac{11}{\sqrt{2}}\right)+33 \sqrt{2}
\mathcal{C}_2-1952 e^{241/4} M^7\right)\, ,
\end{align}
and by simply assuming that $\mathcal{C}_2=\beta/\kappa^7$ and
keeping the leading order terms, we have, regardless the values of
the free parameters $M$, $\beta$, $n_s\sim 0.967078$ which is
quite close to the value $n_s=0.966667$ which corresponds to the
pure $R^2$ model. Now, with regard to the tensor-to-scalar ratio,
for $60$ $e$-foldings and for $\mathcal{C}_2=\beta\kappa^7$ this
reads,
\begin{equation}\label{rapproximatesexpression60}
r\simeq \frac{48 \left(3 \sqrt{2} \beta +33 e^{121/2} \sqrt{\pi }
\beta  \mathrm{Erf}\left(\frac{11}{\sqrt{2}}\right)-176 e^{241/4}
\kappa ^7 M^7\right)^2}{\left(360 \sqrt{2} \beta +3993 e^{121/2}
\sqrt{\pi } \beta  \mathrm{Erf}\left(\frac{11}{\sqrt{2}
}\right)-21296 e^{241/4} \kappa ^7 M^7\right)^2}\, ,
\end{equation}
so by keeping the dominant terms we get at leading order
$r=0.00327846$, and this value is obtained irrespective of the
values of the free parameters $M$ and $\beta$ and it is quite
close to the one corresponding to the pure $R^2$ model which is
$r=0.003333$. Accordingly the tensor spectral index for the model
at hand reads,
\begin{equation}\label{tensorspectralindex}
n_T\simeq \frac{2 e^{241/4} M \left(16 M^7-3 \sqrt[4]{e} \sqrt{\pi
} \mathcal{C}_2
\mathrm{Erf}\left(\frac{11}{\sqrt{2}}\right)\right)}{121 \left(366
e^{121/2} \sqrt{\pi } \mathcal{C}_2
\mathrm{Erf}\left(\frac{11}{\sqrt{2}}\right) M+33 \sqrt{2}
\mathcal{C}_2 M-1952 e^{241/4} M^8\right)}\, ,
\end{equation}
hence by keeping the dominant terms we have approximately
$n_T\simeq -0.000135483$ irrespective of the values of the free
parameters. Thus overall, the pure $R^2$ model and the deformed
model (\ref{frfinal}) have quite similar phenomenology during
inflation. This can also be seen in Fig. \ref{plotplanck} where we
present the combined $(n_s,r)$-plot for the deformed $R^2$ model
confronted with the latest Planck likelihood curves for $N$ chosen
in the range $N=[50,60]$. As it can be seen, the model has quite
elegant viability characteristics comparable and almost identical
with the standard $R^2$ model.
So far we have not defined the integration constant
$\mathcal{C}_2$ which is arbitrary. However, at this point we
shall fix it by using the asymptotic limit of the deformed $f(R)$
gravity (\ref{frfinal}) at late times, thus for $R\to 0$, in which
limit the deformed $f(R)$ reads,
\begin{align}\label{asymptoticinflationfr}
 f(R)\simeq & R+R \left(-\frac{3 \sqrt[4]{e} \sqrt{\pi }
\mathcal{C}_2 \mathrm{Erf}\left(\frac{1}{2}\right)}{16
M^7}-\frac{\mathcal{C}_2}{8 M^7}\right)+\frac{-3 \sqrt[4]{e}
\sqrt{\pi } \mathcal{C}_2 M
\mathrm{Erf}\left(\frac{1}{2}\right)-18 \mathcal{C}_2
\sqrt{M^2}+16 M^8}{96 M^6}+\frac{R^2}{6 M^2}\\ \notag & +\frac{R^2
\left(-3 \sqrt[4]{e} \sqrt{\pi } \mathcal{C}_2 M
\mathrm{Erf}\left(\frac{1}{2}\right)-6 \mathcal{C}_2 M\right)}{96
M^{10}}\, ,
\end{align}
and we can fix $\mathcal{C}_2$ by fixing the asymptotical limit
(\ref{asymptoticinflationfr}) to have zero effective cosmological
constant. Thus if we demand for late-time reasoning the last term
to vanish, the constant $\mathcal{C}_2$ must be chosen as
$\mathcal{C}_2=\frac{16 M^7}{3 \left(\sqrt[4]{e} \sqrt{\pi }
\mathrm{Erf}\left(\frac{1}{2}\right)+6\right)}$ so the asymptotic
form of the $f(R)$ gravity (\ref{asymptoticinflationfr}) becomes,
\begin{align}\label{asymptoticformfrzero}
 f(R)\sim & R+R\left(-\frac{\sqrt[4]{e} \sqrt{\pi }
\mathrm{Erf}\left(\frac{1}{2}\right)}{\sqrt[4]{e} \sqrt{\pi }
\mathrm{Erf}\left(\frac{1}{2}\right)+6}-\frac{2}{3
\left(\sqrt[4]{e} \sqrt{\pi }
\mathrm{Erf}\left(\frac{1}{2}\right)+6\right)}\right)+\frac{R^2
\left(-\frac{16 \sqrt[4]{e} \sqrt{\pi }
\mathrm{Erf}\left(\frac{1}{2}\right) M^8}{\sqrt[4]{e} \sqrt{\pi }
\mathrm{Erf}\left(\frac{1}{2}\right)+6}-\frac{32 M^8}{\sqrt[4]{e}
\sqrt{\pi } \mathrm{Erf}\left(\frac{1}{2}\right)+6}+16
M^8\right)}{96 M^{10}}\, .
\end{align}
It is notable that the late-time era for the model at hand is
described by a rescaled Einstein-Hilbert term. This is an
effective theory with a weaker gravity compared to the
Einstein-Hilbert term. Basically, the leading order
Einstein-Hilbert term is a rescaled Einstein-Hilbert term with a
smaller effective Planck mass compared to the simple
Einstein-Hilbert term $R$. In the next section, we shall propose
an effective correction term which has no effect during the
inflationary era, but when it is combined with the $f(R)$ gravity
(\ref{asymptoticformfrzero}), results to an early dark energy
evolution at late times, and more importantly, it results to a
viable dark energy era at present time.

\section{Late-time Behavior and Proposal for Dark Energy Correction Terms}

In this section we shall investigate the late-time perspectives of
the deformed $R^2$ model of Eq. (\ref{frfinal}) by also adding a
correction term of the form $R\,e^{26\Lambda/R}e^{-R/M^2}$ where
$\Lambda$ is the present time cosmological constant,
$\Lambda\simeq 11.895\times 10^{-67}$eV$^2$ and $M$ is the
parameter appearing in the deformed $R^2$ model, which in for
strictly phenomenological reasons related to the amplitude of the
scalar perturbations in the Einstein frame version of the pure
$R^2$ model, $M$ must be chosen $M= 1.5\times
10^{-5}\left(\frac{N}{50}\right)^{-1}M_p$ \cite{Appleby:2009uf},
hence for $N\sim 60$, $M$ is $M\simeq 3.04375\times 10^{22}$eV. At
late-time thus, by using the expansion
(\ref{asymptoticformfrzero}), and by adding the correction term
$R\,e^{26\Lambda/R}e^{-R/M^2}$, the late-time $f(R)$ gravity
reads,
\begin{align}\label{frlatetimefinal}
f(R)\simeq  & R+\frac{\left(-\frac{16 \sqrt[4]{e} \sqrt{\pi }
\mathrm{Erf}\left(\frac{1}{2}\right)}{\sqrt[4]{e} \sqrt{\pi }
\mathrm{Erf}\left(\frac{1}{2}\right)+6}-\frac{32}{\sqrt[4]{e}
\sqrt{\pi } \mathrm{Erf}\left(\frac{1}{2}\right)+6}+16\right)
R^2}{96 M^2}\\ \notag &+\left(-\frac{\sqrt[4]{e} \sqrt{\pi }
\mathrm{Erf}\left(\frac{1}{2}\right)}{\sqrt[4]{e} \sqrt{\pi }
\mathrm{Erf}\left(\frac{1}{2}\right)+6}-\frac{2}{3
\left(\sqrt[4]{e} \sqrt{\pi }
\mathrm{Erf}\left(\frac{1}{2}\right)+6\right)}\right) R+R \exp
\left(-\frac{R}{M^2}\right) \exp \left(\frac{26 \Lambda
}{R}\right)\, .
\end{align}
Note that the unified description scheme of inflation with dark
energy in the context of $f(R)$ gravity, was first proposed in
\cite{Nojiri:2003ft} and extended to a number of models in the
review \cite{reviews4}. Also, one can consider such unification of
deformed $R^2$ inflation with dark energy, using other $f(R)$
gravity terms which do not affect inflation, but produce a viable
dark energy era, such as subdominant power-law terms. A couple of
comments prior to continuing to the phenomenological analysis of
the model at hand. Firstly, the correction term
$R\,e^{26\Lambda/R}e^{-R/M^2}$ has no effect during early times,
when the curvature takes large values. Thus it is essentially an
early dark energy term which is inactive at early times but takes
over the control of the dynamics of the model at late times. This
issue is important and at this point we shall analyze it in
detail. One question is, why did not the early dark energy
correction term $\sim R\,e^{26\Lambda/R}e^{-R/M^2}$ did not appear
from the first place in the $f(R)$ gravity which realizes the
quasi-de Sitter evolution, namely in Eq. (\ref{frfinal}). The
answer is simple, the $f(R)$ gravity in Eq. (\ref{frfinal}) is the
$f(R)$ gravity which realizes the quasi-de Sitter patch of the
Universe at early times. Hence it is a leading order behavior of
the full underlying $f(R)$ gravity, which is unknown to us. Hence
in general, there might be many other subdominant terms during the
quasi-de Sitter era of the Universe, which are unknown to us. One
of these possible terms could be the one we introduced $\sim
R\,e^{26\Lambda/R}e^{-R/M^2}$, which is truly subdominant during
inflation, as we now evince. It is a big problem for scientists to
know the exact form of $f(R)$ gravity which may control the full
evolution history of our Universe, and the problem relies to our
inability to find a totally unified evolution of the Universe in
terms of a unique Hubble rate. Instead of that, we have only clues
for four distinct evolutionary era of our Universe, the
inflationary era, usually realized successfully by a quasi-de
Sitter evolution, the radiation domination era followed by a
matter domination era which lastly is followed by a dark energy
era. Thus we basically have patches of evolution for which we know
some things and not a unified evolutionary behavior in terms of a
unique Hubble rate. An example of a unified scalar factor
describing all the eras would be,
\begin{equation}\label{aunified}
a(t)=a_0e^{\mathcal{H}_0 t-\frac{1}{12} M^2
t^2}+a_et^{1/2}+a_It^{2/3}+a_{II}e^{\sqrt{\Lambda}
t-\Lambda_1t^2}\, ,
\end{equation}
where $a_0$ denotes the scale factor at the beginning of the
inflationary era, $a_e$ denotes the scale factor at the end of
inflation and at the start of the radiation domination era,
$a_{I}$  denotes the scale factor at the end of the radiation
domination era and at the beginning of the dark matter domination
era, and finally $a_{II}$ denotes the scale factor at the end of
the matter domination era and at the beginning of the dark energy
era. As it is apparent from Eq. (\ref{aunified}), it is impossible
to find using the reconstruction technique we introduced in the
previous section which $f(R)$ gravity realizes such an evolution,
at least in an analytic way. It is simply an inevitable task.
However what we certainly do is to find which underlying $f(R)$
gravity may realize each evolutionary patch separately. Thus in
the previous section we showed that the dominant $f(R)$ gravity
form which is responsible for the realization of the quasi-de
Sitter patch, namely the first term in Eq. (\ref{aunified}), is
given by the $f(R)$ gravity appearing in Eq. (\ref{frfinal}). It
is however highly possible that many subdominant terms appear in
the effective inflationary Lagrangian, which are responsible for
the realization of one of the three last terms in Eq.
(\ref{aunified}). An example of this sort is played by the early
dark energy term which we added by hand, namely $\sim
R\,e^{26\Lambda/R}e^{-R/M^2}$. This term is totally ignorable
during inflation, and is actually activated only at late times,
and specifically for the redshift range $z=[0,10]$. Let us show
that this term is indeed inactive during inflation. As we
mentioned in the beginning of this section, for the dark energy
era analysis we shall take $\Lambda\simeq 11.895\times
10^{-67}$eV$^2$ and $M\simeq 3.04375\times 10^{22}$eV and also the
scale of inflation is $H_I\sim 10^{16}$GeV. For these values, and
during inflation, in which case $R\sim 12H_I^2$, the term $\sim
\frac{R^2}{6 M^2}$ is of the order $\frac{R^2}{6 M^2}\sim
\mathcal{O}(10^{56})$eV, while the term $R\sim
\mathcal{O}(10^{50})$eV and the early dark energy term
$R\,e^{26\Lambda/R}e^{-R/M^2}\sim \mathcal{O}(10^{-562482})$eV.
Thus it is apparent that the early dark energy term is highly
subdominant during inflation and therefore it does not affect at
all the dynamics during inflation. However as we evince in this
section, it affects significantly the late time dynamics of the
cosmological system.

\begin{figure}
\centering
\includegraphics[width=18pc]{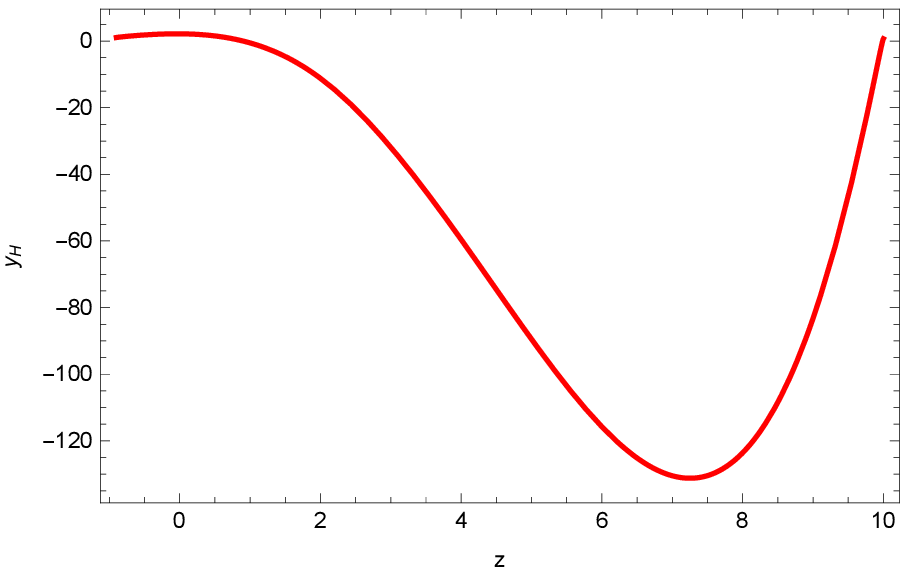}
\includegraphics[width=18pc]{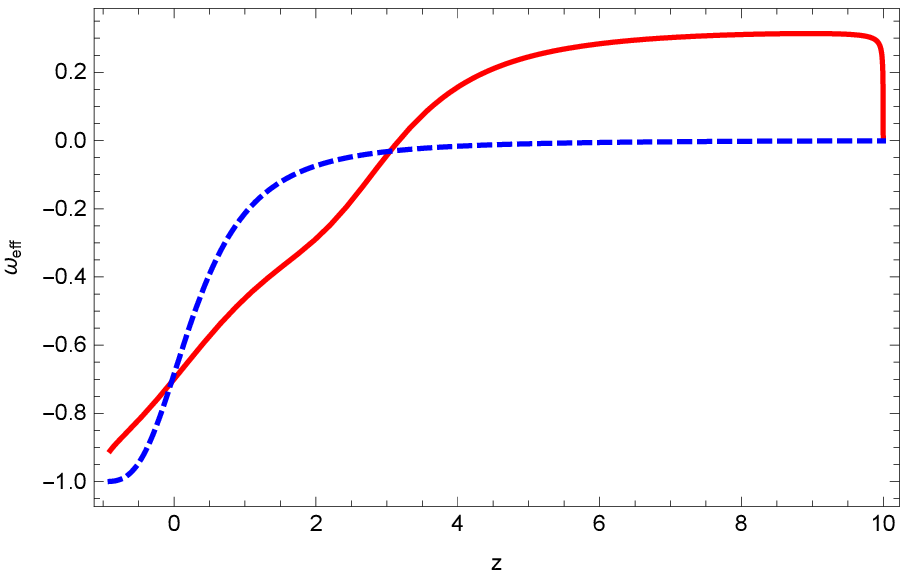}
\includegraphics[width=18pc]{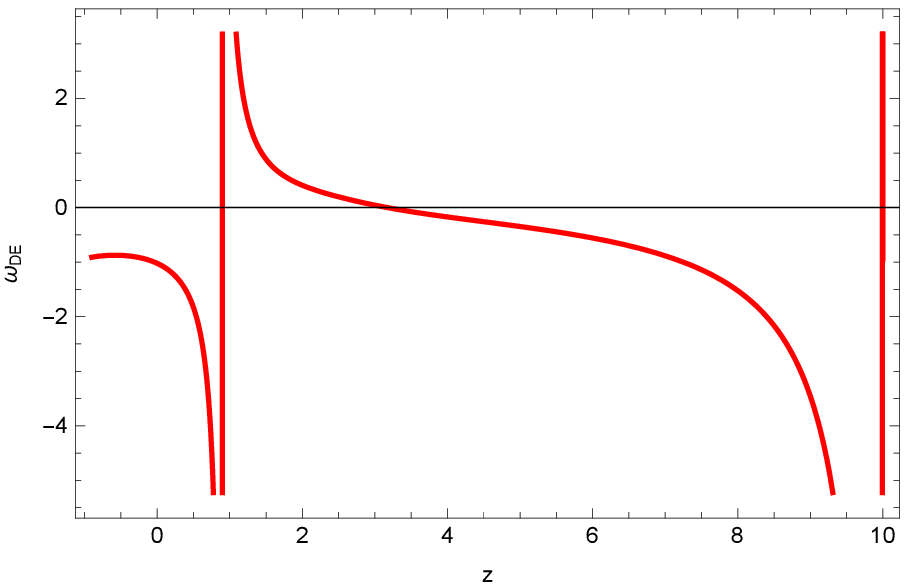}
\caption{The function $y_H$ (left plot), the total EoS parameter
$\omega_{tot}$ (right plot) and the dark energy EoS parameter
$\omega_{DE}(z)$ (bottom plot) as functions of the redshift. for
The deformed $R^2$ model corresponds to red curves and the
$\Lambda$CDM model to blue curves.}\label{plot3}
\end{figure}

\begin{figure}
\centering
\includegraphics[width=18pc]{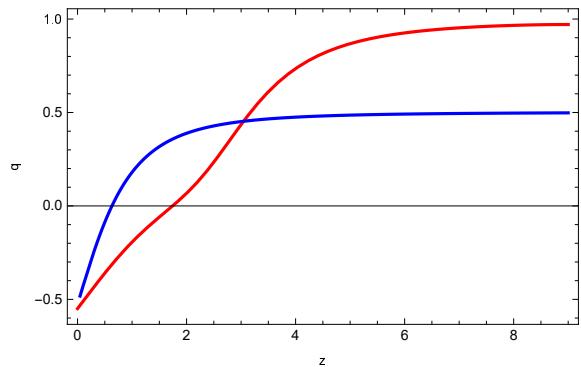}
\caption{The deceleration parameter, as a function of the redshift
for the deformed $R^2$ model (red curve) and for the $\Lambda$CDM
model (blue curve).} \label{plot4}
\end{figure}

\begin{figure}[h!]
\centering
\includegraphics[width=18pc]{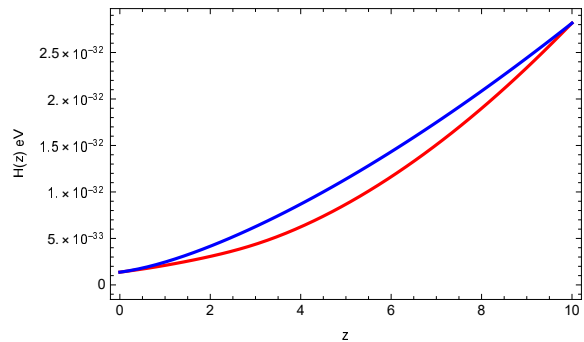}
\caption{The Hubble rate $H(z)$ as a function of the redshift for
the deformed $R^2$ model (red curve) and for the $\Lambda$CDM
model (blue curve).} \label{plot5}
\end{figure}

Furthermore, the model (\ref{frlatetimefinal}) has effectively a
larger effective Newton gravitational constant, or a smaller
effective Planck mass, compared to Einstein-Hilbert gravity. This
is due to the fact that the deformed $R^2$ gravity of Eq.
(\ref{frfinal}) asymptotically for small curvature values at
linear order in $R$ behaves as,
\begin{equation}\label{frorderone}
f(R)\sim R+R\left(-\frac{\sqrt[4]{e} \sqrt{\pi }
\mathrm{Erf}\left(\frac{1}{2}\right)}{\sqrt[4]{e} \sqrt{\pi }
\mathrm{Erf}\left(\frac{1}{2}\right)+6}-\frac{2}{3
\left(\sqrt[4]{e} \sqrt{\pi }
\mathrm{Erf}\left(\frac{1}{2}\right)+6\right)}\right) \, ,
\end{equation}
and the second term is basically a negative term which effectively
reduces the effective Planck mass or equivalently it increases the
strength of gravity at late times. Hence for the model at hand,
the late-time era gravity is stronger at linear order, compared to
the standard Einstein-Hilbert term. Now we shall analyze in detail
the late-time dynamics of the model (\ref{frlatetimefinal}),
emphasizing in studying quantities of cosmological interest which
we now discuss. We shall assume that along with the $f(R)$ gravity
model (\ref{frlatetimefinal}), radiation and cold dark matter
fluids are present, which were neglected during the early time
era. For the $f(R)$ gravity in the presence of matter fluids, the
field equations can be written in the Einstein-Hilbert form in the
following way,
\begin{align}\label{flat}
& 3H^2=\kappa^2\rho_{tot}\, ,\\ \notag &
-2\dot{H}=\kappa^2(\rho_{tot}+P_{tot})\, ,
\end{align}
where the total energy density is equal to
$\rho_{tot}=\rho_{m}+\rho_{DE}+\rho_r$, and $\rho_m$ and $\rho_r$
are the matter and radiation fluids energy densities respectively.
Also the dark energy density $\rho_{DE}$ is the geometric
contribution of the $f(R)$ gravity fluid, which essentially will
control the late-time dynamics, and it is defined as follows,
\begin{equation}\label{degeometricfluid}
\kappa^2\rho_{DE}=\frac{f_R R-f}{2}+3H^2(1-f_R)-3H\dot{f}_R\, .
\end{equation}
The corresponding total pressure $P_{tot}=P_r+P_{DE}$, and the
dark energy pressure is defined as,
\begin{equation}\label{pressuregeometry}
\kappa^2P_{DE}=\ddot{f}_R-H\dot{f}_R+2\dot{H}(f_R-1)-\kappa^2\rho_{DE}\,
.
\end{equation}
In order to study the late-time dynamics of the model
(\ref{frlatetimefinal}), we shall introduce some statefinder
quantities in terms of the redshift, which we shall use as a
dynamical parameter. This will facilitate our late-time study.
Firstly, the redshift $z$ parameter is,
\begin{equation}\label{redshift}
1+z=\frac{1}{a}\, ,
\end{equation}
and we assumed that present time scale factor, at $z=0$, is equal
to unity. In this way, physical wavelengths and comoving
wavelengths coincide at late times. We introduce the statefinder
function $y_H(z)$
\cite{Hu:2007nk,Bamba:2012qi,Odintsov:2020vjb,Odintsov:2020nwm,Odintsov:2020qyw,reviews1},
\begin{equation}\label{yHdefinition}
y_H(z)=\frac{\rho_{DE}}{\rho^{(0)}_m}\, ,
\end{equation}
where $\rho^{(0)}_m$ is the energy density of cold dark matter at
present time. Being a statefinder quantity, $y_H(z)$ quantifies
the effects of dark energy at late times, and it is certainly
different from zero, when geometric terms affect the late-time
dynamics. This can be true in two cases, firstly when a
cosmological constant is directly present in the effective
Lagrangian, or even when geometric terms of a higher order gravity
are present in the Lagrangian. If $y_H$ is zero, this case simply
describes the Einstein-Hilbert FRW cosmology without a
cosmological constant. Using Eq. (\ref{flat}), we can write the
function $y_H(z)$ as follows,
\begin{equation}\label{finalexpressionyHz}
y_H(z)=\frac{H^2}{m_s^2}-(1+z)^{3}-\chi (1+z)^4\, ,
\end{equation}
where recall that $\rho_m=\rho^{(0)}_m (1+z)^3$ and also we
introduced $\chi$ defined as
$\chi=\frac{\rho^{(0)}_r}{\rho^{(0)}_m}\simeq 3.1\times 10^{-4}$,
and $\rho^{(0)}_r$ denotes the radiation energy density at present
time. Also the parameter $m_s$ is defined as
$m_s^2=\frac{\kappa^2\rho^{(0)}_m}{3}=H_0\Omega_c=1.37201\times
10^{-67}$eV$^2$, and we used the Planck 2018 constraints on the
cosmological parameters, which indicate that $H_0\simeq
1.37187\times 10^{-33}$eV. Hence the statefinder quantity $y_H(z)$
indicates deviations from the standard $\Lambda$CDM model, in
which case it would simply be a constant. In the case of $f(R)$
gravity, the function $y_H(z)$ indicates the dynamical dark energy
character of the late-time behavior. Rewriting the Friedmann
equations in terms of the function $y_H(z)$ and using the redshift
parameter as a dynamical parameter we get,
\begin{equation}\label{differentialequationmain}
\frac{d^2y_H(z)}{d z^2}+J_1\frac{d y_H(z)}{d z}+J_2y_H(z)+J_3=0\,
,
\end{equation}
where the dimensionless functions $J_1$, $J_2$ and $J_3$ are
defined as,
\begin{align}\label{diffequation}
& J_1=\frac{1}{z+1}\left(
-3-\frac{1-F_R}{\left(y_H(z)+(z+1)^3+\chi (1+z)^4\right) 6
m_s^2F_{RR}} \right)\, , \\ \notag & J_2=\frac{1}{(z+1)^2}\left(
\frac{2-F_R}{\left(y_H(z)+(z+1)^3+\chi (1+z)^4\right) 3
m_s^2F_{RR}} \right)\, ,\\ \notag & J_3=-3(z+1)-\frac{\left(1-F_R
\right)\Big{(}(z+1)^3+2\chi (1+z)^4
\Big{)}+\frac{R-F}{3m_s^2}}{(1+z)^2\Big{(}y_H(z)+(1+z)^3+\chi(1+z)^4\Big{)}6m_s^2F_{RR}}\,
,
\end{align}
where we defined $f(R)=R+F(R)$ and $F_{RR}=\frac{\partial^2
F}{\partial R^2}$, while $F_{R}=\frac{\partial F}{\partial R}$. We
shall solve numerically the differential equation
(\ref{differentialequationmain}) focusing on the late-time
redshift interval $z=[0,10]$, by using following set of initial
conditions
\begin{equation}\label{generalinitialconditions}
y_H(z_f)=\frac{\Lambda}{3m_s^2}\left(
1+\frac{(1+z_f)}{1000}\right)\, , \,\,\,\frac{d y_H(z)}{d
z}\Big{|}_{z=z_f}=\frac{1}{1000}\frac{\Lambda}{3m_s^2}\, ,
\end{equation}
which are well motivated and justified by the late moments of the
matter domination era
\cite{Hu:2007nk,Bamba:2012qi,Odintsov:2020vjb,Odintsov:2020nwm,Odintsov:2020qyw,reviews1}.
For our analysis we shall focus on several physical cosmology
quantities of interest, which we shall rewrite in terms of the
statefinder function $y_H(z)$. A vital quantity is the curvature,
which in terms of $y_H(z)$ it is written as,
\begin{equation}\label{ricciscalarasfunctionofz}
R(z)=3m_s^2\left( 4y_H(z)-(z+1)\frac{d y_H(z)}{d
z}+(z+1)^3\right)\, .
\end{equation}
Accordingly, the dark energy density parameter $\Omega_{DE}$ is
written as,
\begin{equation}\label{omegaglarge}
\Omega_{DE}(z)=\frac{y_H(z)}{y_H(z)+(z+1)^3+\chi (z+1)^4}\, ,
\end{equation}
and this quantity is quite important since its present day value
is accurately constrained by the Planck 2018 constraints
\cite{Planck:2018vyg}. The same applies for the dark energy EoS
parameter $\omega_{DE}$, which is,
\begin{equation}\label{omegade}
\omega_{DE}(z)=-1+\frac{1}{3}(z+1)\frac{1}{y_H(z)}\frac{d
y_H(z)}{d z}\, ,
\end{equation}
and a parameter which is quite important for the overall behavior
of the model is the total EoS parameter which reads,
\begin{equation}\label{totaleosparameter}
\omega_{tot}(z)=\frac{2 (z+1) H'(z)}{3 H(z)}-1\, .
\end{equation}
With regard to popular statefinder quantities, we shall choose the
deceleration parameter $q$, which in terms of the redshift is
defined as follows,
\begin{align}\label{statefinders}
& q=-1-\frac{\dot{H}}{H^2}=-1+(z+1)\frac{H'(z)}{H(z)}\, .
\end{align}
In the following we shall compare the dark energy corrected model
deformed $R^2$ model (\ref{frlatetimefinal}) with the Planck data
at present time, or the $\Lambda$CDM model where it is possible,
with the  $\Lambda$CDM model Hubble rate in terms of the redshift
being defined as follows,
\begin{equation}\label{lambdacdmhubblerate}
H_{\Lambda}(z)=H_0\sqrt{\Omega_{\Lambda}+\Omega_M(z+1)^3+\Omega_r(1+z)^4}\,
,
\end{equation}
where $H_0$ is the value of the Hubble rate at present day.
Moreover, $\Omega_{\Lambda}\simeq 0.681369$ and $\Omega_M\sim
0.3153$ \cite{Planck:2018vyg}, while $\Omega_r/\Omega_M\simeq
\chi$, and recall $\chi$ is defined below Eq.
(\ref{finalexpressionyHz}).
Let us present now the results of our numerical analysis in some
detail. In Fig. \ref{plot3} we present the behavior of the
function $y_H(z)$ (left plot), the total EoS parameter
$\omega_{tot}$ (right plot) and the dark energy EoS parameter
(bottom plot) as functions of the redshift. The behavior of $y_H$
indicates firstly that the model deviates from the $\Lambda$CDM
model and it basically describes an early dark energy era. The
same conclusion can be obtained by looking at the plot of the dark
energy EoS parameter, where two steep acceleration eras appear to
have occurred in the past. The total EoS parameter in the right
plot of Fig. \ref{plot3} contains the behavior of the total EoS
parameter for the deformed $R^2$ model (red curve) compared with
the $\Lambda$CDM plot (blue curve). One thing is apparent for
sure, qualitatively the deformed $R^2$ model behaves as the
$\Lambda$CDM model, however it seems that the deformed $R^2$ model
is a deformation of the $\Lambda$CDM model and there are clear
distinctions between the two models.
The same conclusion can be reached by looking Fig. \ref{plot4}
where we plot the deceleration parameter $q$ for the deformed
$R^2$ model (red curves) and for the $\Lambda$CDM model (blue
curves). A notable feature for the deformed $R^2$ model is that
the transition from deceleration to acceleration occurs earlier
compared to the $\Lambda$CDM model. Finally, in Fig. \ref{plot5}
we plot the Hubble rate for for the deformed $R^2$ model (red
curves) and for the $\Lambda$CDM model (blue curves). In general
the $\Lambda$CDM model has larger values for the Hubble rate
during the whole range $z=[0,10]$ however, at present day the
deformed $R^2$ model has slightly elevated value for the Hubble
rate compared to the $\Lambda$CDM model. In Table \ref{table1} we
gather some characteristic values for the several physical
cosmology quantities of interest. As it can be seen, the dark
energy deformed $R^2$ model is quite compatible with the latest
Planck constraints on cosmological parameters
\cite{Planck:2018vyg}. One of the most important parameters is the
dark energy EoS parameter, which for the deformed $R^2$ model it
reads $\omega_{DE}(0)\simeq -1.02159$ which lies within the
viability limits of the Planck constraint $\omega_{DE}=-1.018\pm
0.031$. A notable feature, mostly absent in the majority of $f(R)$
gravity models is the fact that the dark energy era at present day
is slightly phantom. Another important quantity is the dark energy
density parameter, which for the deformed $R^2$ model reads,
$\Omega_{DE}(0)\simeq 0.684831$, and the Planck constraints on
this are $\Omega_{DE}=0.6847\pm 0.0073$. Also in Table
\ref{table1} it can be seen that the Hubble rate value at present
day is slightly elevated for the deformed $R^2$ model in
comparison to the $\Lambda$CDM model, and it is contrasted with
the overall behavior of the two models for intermediate redshifts.
\begin{table}[h!]
  \begin{center}
    \caption{\emph{\textbf{Values of Cosmological Parameters for Deformed $R^2$ Gravity Model and $\Lambda$CDM Model }}}
    \label{table1}
    \begin{tabular}{|r|r|r|}
     \hline
      \textbf{Cosmological Parameter} & \textbf{Deformed $R^2$ Gravity Value} & \textbf{Base $\Lambda$CDM or Planck 2018 Value} \\
           \hline
      $\Omega_{DE}(0)$ & 0.684831 & $0.6847\pm 0.0073$ \\ \hline
      $\omega_{DE}(0)$ & -1.02159 & $-1.018\pm 0.031$\\ \hline
      $q(0)$ & -0.54938 & -0.535\\
      \hline
      $H_0$ in eV & $1.37237\times 10^{-33}$ & $1.37187\times 10^{-33}$\\
      \hline
    \end{tabular}
  \end{center}
\end{table}
Overall, the dark energy corrected deformed $R^2$ model provides a
viable description for the dark energy era, which mimics
qualitatively the evolution of the $\Lambda$CDM model, without
overlaps between the two models though. Thus with the extra dark
energy correction term, which during inflation does not affect at
all the dynamics, but takes control over the dynamics of the model
at late times, it is possible to provide a unified description of
the inflationary era and the dark energy era with the same $f(R)$
gravity model.

\section{Conclusions}

In this work we pointed out that a quasi-de Sitter evolution in
$f(R)$ gravity is not solely realized by the vacuum $R^2$ model,
as it is widely known in the literature, but from a deformed
$f(R)$ gravity, containing the $R^2$ model. By using a standard
reconstruction technique, we calculated the exact form of the
$f(R)$ gravity which generates an exact quasi-de Sitter evolution.
For the $f(R)$ gravity we found, we calculated in detail the
slow-roll indices and the observational indices of inflation,
focusing on the spectral index of the primordial scalar and tensor
perturbations, and the tensor-to-scalar ratio. As we demonstrated,
the resulting model generates and inflationary evolution which is
quantitatively similar to the $R^2$ model. The deformed $R^2$
model which generates the exact quasi-de Sitter evolution, at
late-times yields an interesting characteristic, a stronger
gravity in terms of the effective Newton's constant at late times,
or equivalently a smaller effective Planck mass at late times. We
introduced an effective early dark energy correction term, which
although did not affect the inflationary era, it affects the
late-time evolution. Particularly, the dark energy era predicted
by the deformed $R^2$ model we found is a deformed version of the
$\Lambda$CDM model, which is viable when confronted with the
Planck 2018 data at present day, concerning the dark energy EoS
and density parameters. Thus with this work, we revealed a hidden
aspect of an inflationary quasi-de Sitter evolution in the context
of $f(R)$ gravity and we showed that there are effective
correction terms to the standard $R^2$ model, which although they
slightly affect the early time era, these may affect significantly
the late-time era. The importance of the quasi-de Sitter evolution
compels to perform further realizations of this important
evolution patch of the Universe in other modified gravity
frameworks, such as pure extended Gauss-Bonnet gravity
\cite{Oikonomou:2015qha} or even Einstein-Gauss-Bonnet gravity
\cite{Koh:2014bka,Odintsov:2020sqy}, or extensions of these
theories, like Horndeski theories \cite{Bayarsaikhan:2020jww}. We
hope to address these issues in future works.

\section*{Acknowledgments}

This work was supported by MINECO (Spain), project
PID2019-104397GB-I00 (S.D.O).

\end{document}